\documentclass[twocolumn]{aastex701}

\usepackage{graphicx}

\received{26 August 2025}
\revised{12 September 2025}
\accepted{18 September 2025}

\shorttitle{Caught in the Cosmic Web}
\shortauthors{Luber et al.}

\begin{document}

\title{Caught in the Cosmic Web: Evidence for Ram-Pressure Stripping of a Low-Mass Galaxy by the Cosmic Web}

\correspondingauthor{Nick Luber}
\email{nicholas.m.luber@gmail.com}

\author[orcid=0009-0006-6641-0928]{Nicholas Luber}
\affiliation{Department of Astronomy, Columbia University, Mail Code 5247, 538 West 120th Street, New York, NY 10027, USA}
\email{nicholas.m.luber@gmail.com}

\author[orcid=0000-0002-7532-3328]{Sabrina Stierwalt}
\affiliation{Department of Physics, Occidental College, 1600 Campus Road, Los Angeles, CA 90041, USA}
\email{sabrina@oxy.edu} 

\author[orcid=0000-0003-3474-1125]{George C. Privon}
\affiliation{National Radio Astronomy Observatory, 520 Edgemont Road, Charlottesville, VA 22903, USA}
\affiliation{Department of Astronomy, University of Virginia, 530 McCormick Road, Charlottesville, VA 22904 USA}
\affiliation{Department of Astronomy, University of Florida, P.O. Box 112055, 
Gainesville, FL 32611, USA}
\email{gprivon@nrao.edu}

\author[0000-0003-0715-2173]{Gurtina Besla}
\affiliation{Department of Astronomy, University of Arizona, 933 North Cherry Avenue, Tucson, AZ 85721}
\email{gbesla@email.arizona.edu}

\author[0000-0001-8348-2671]{Kelsey E. Johnson}
\affiliation{Department of Astronomy, University of Virginia, 530 McCormick Road, Charlottesville, VA 22904 USA}
\email{kej7a@virginia.edu}

\author[0000-0002-3204-1742]{Nitya Kallivayalil}
\affiliation{Department of Astronomy, University of Virginia, 530 McCormick Road, Charlottesville, VA 22904 USA}
\email{njk3r@virginia.edu}

\author[orcid=0000-0002-1871-4154]{David R. Patton}
\affiliation{Department of Physics \& Astronomy, Trent University, 1600 West Bank Drive, Peterborough, Ontario, Canada K9L OG2}
\email{dpatton@trentu.ca}

\author[0000-0002-1129-1873]{Mary Putman}
\affiliation{Department of Astronomy, Columbia University, Mail Code 5247, 538 West 120th Street, New York, NY 10027, USA}
\email{mep2157@columbia.edu}

\author[0000-0002-9001-6713]{Jingyao Zhu}
\affiliation{Department of Astronomy, Columbia University, Mail Code 5247, 538 West 120th Street, New York, NY 10027, USA}
\email{mep2157@columbia.edu}

\begin{abstract}

We present interferometric radio observations of the neutral atomic gas in AGC 727130, a low-mass, gas-rich, field galaxy lacking significant star-formation. The atomic gas in AGC 727130 displays a pronounced asymmetry, extending well beyond the stellar disk in one direction while remaining relatively undisturbed in the other. Despite proximity to a pair of interacting dwarfs, tidal analysis suggests these neighboring galaxies are not responsible for this pronounced asymmetry. Instead, using a topological cosmic web filament finder on spectroscopic catalogue data, we find AGC 727130 lies at the intersection of several large-scale cosmic web filaments, environments predicted to host diffuse, shock-heated gas. We propose that an interaction with this ambient medium is stripping gas from the galaxy via cosmic web ram-pressure stripping. This mechanism, supported by recent simulations, may quench low-mass galaxies outside of massive halos, and must be accounted for when comparing observed numbers of dwarf galaxies to theoretical predictions.

\end{abstract}


\section{Introduction}\label{sec:Introduction}

\quad The Lambda Cold Dark Matter \citep[LCDM,][]{White78, Blumenthall84} paradigm for galaxy evolution has been remarkably successful in describing the large-scale structure (LSS) of the universe, accurately reproducing the cosmic web, the distribution of galaxy clusters, and the statistical properties of large-scale matter fluctuations. However, despite these successes, it struggles to reproduce key observations in the low mass galaxy regime \citep{Bullock17}. In particular, uncertainty surrounding which mechanisms dominate quenching in dwarf galaxies has produced an evolving set of differences between observations and LCDM predictions. The ``missing satellites problem'' was first introduced when simulations of the formation of a Milky Way-mass host consistently produced an order of magnitude more low mass halos than the observed number of dwarf galaxies in the Local Group \citep{Klypin99, Moore99}. This discrepancy has largely been resolved in part due to a better understanding of baryonic physics within the dwarf galaxies themselves, including quenching mechanisms that can prevent star formation. Photoionization during the reionization era, for example, heated the intergalactic medium and prevented low-mass halos from accreting and cooling gas, thereby stifling early star formation \citep{Thoul96, Quinn96, Navarro97}. 

\quad More recently studies of the fraction of quenched dwarf satellites around massive galaxy hosts at first showed discrepancies between observations and simulations when only higher mass dwarfs were included. Simulations typically produce quenched fractions of $\sim$70\% \citep{Font21, Karunakaran21}, similar to what is observed in the Local Group \citep{Wetzel15}, while observed satellite mass functions around other Milky Way-like hosts probing satellites down to roughly the mass of the SMC showed quenched fractions of at most 20\% \citep{SAGA1, SAGA2}. However, as techniques for finding lower and lower mass satellites have improved, fractions of quenched dwarfs as high as 80\% are now observed in studies that include dwarfs down to M$_*$ $\sim$ 10$^7$M$_{\odot}$, in good agreement with $\Lambda$-CDM \citep{elves, saga3,saga4}.  

\quad Adding to the uncertainty as to what causes quenching in dwarf galaxies is the fact that low mass, quenched galaxies are extremely rare in the field. Less than 0.06\% of galaxies with 10$^7$ M$_{\odot}$ $\lesssim$ M$_*$ $\lesssim$ 10$^9$ M$_{\odot}$ are observed to be quenched outside the presence of a massive host \citep{Geha12}. Interactions between dwarf galaxies are further found to not be a reliable mechanism for turning off star formation \citep{Stierwalt15}. With only a few quenched dwarf galaxies known to exist in the field, their quenching mechanisms are not always clear \citep{KF24}. Ram pressure stripping (RPS), caused by a galaxy moving through a dense medium, such as a galaxy cluster \citep{GunnGott}, can strip away a galaxy's interstellar gas, particularly its atomic neutral hydrogen (HI), effectively starving it of fuel for future star formation \citep{Kenney04, Zhu24, Zabel24}. Importantly, unlike tidal interactions which perturb both stars and gas, RPS is a hydrodynamic process that acts solely on the gaseous component while leaving the stellar disk largely undisturbed providing a clear diagnostic for identifying RPS in action \citep{Boselli22}. While RPS is typically invoked to explain quenching in dense environments, its effectiveness far from massive hosts or an intracluster medium remains uncertain. In such contexts, traditional environmental mechanisms fall short, raising the possibility that large-scale structure of the universe itself may contribute to gas removal.

\quad An example of one such process is cosmic web ram-pressure stripping (CWS), a mechanism in which the diffuse gas reservoirs of low-mass galaxies are removed by the movement of galaxies through the large-scale structure of the universe. Recent observational work has begun to disentangle the effects of large-scale structure versus those due to the more local environment on the HI content of individual galaxies in the local universe (z $<$ 0.05) \citep{Kleiner17, Odekon18, Hoosain24}. However, these studies have not converged upon a consensus, with some finding an HI enrichment close to LSS filaments for the most massive galaxies ($M_{\*}$ $>$ 10$^{11}$M$_{\odot}$) \citep{Kleiner17}, and others finding an HI deficiency for galaxies residing near filaments \citep{Odekon18, Hoosain24}. The difference in results for these various works could be attributed to selection biases, LSS reconstruction methods, and different ways of binning the data, i.e. stellar mass, cosmic web distance, color, etc. At intermediate redshifts (z$\approx$0.35), similar studies have been undertaken, but utilizing HI stacking techniques to probe average HI content, as opposed to the use of direct detections \citep{Sinigaglia24, Luber25}. These works find that galaxies closer to filaments are HI rich, in line with the results from \citet{Kleiner17}. 

\quad Focusing on the study of individual galaxies, in a specific case study of Wolf–Lundmark–Melotte (WLM), an irregular dwarf galaxy on the edge of the Local Group, \citet{Yang22} speculate that the presence of HI clouds removed from WLM is due to ram-pressure stripping by a reservoir of intergalactic gas not associated with any local major galaxy, though the results are not definitive. Additionally, using observations of HI, H$\alpha$, and CO, the Virgo Filament Survey find that filaments act as transitional environments where gas depletion, morphological changes, and star formation quenching are already underway before galaxies reach Virgo. These data show that multiple processes, including RPS and tidal interactions, disrupt the baryon cycle in filament and group galaxies, shaping their gas content and star formation histories \citep{Castignani22a,Castignani22b,Finn25}. In conjunction with these observationally motivated studies, recent theoretical work has shown that CWS can effectively remove the gas of low-mass galaxies and quench star-formation for galaxies in the current epoch \citep{Keres09, Thompson23, Pasha23, Benavides25}. The confluence of results from observations and simulations suggest that a galaxy's placement within, and interaction with, the large-scale structure may have an important role in its evolution. Specifically, the CWS by the filaments and nodes of the cosmic web may prevent dwarf galaxies from retaining their gas, effectively starving them before they can evolve into intermediate-to-high-mass galaxies. 

\quad In this work, we explore the case of AGC 727130, a gas-rich, low-mass field galaxy that exhibits signs of CWS and lacks significant star-formation. By analyzing its environment, morphology, and kinematics, we assess the viability of cosmic web stripping as a natural explanation for its characteristics, and more broadly, how this process fits within in galaxy evolution paradigms. In this work, we assume a standard flat FLRW cosmology with H$_{0}$ = 70 km s$^{-1}$ Mpc$^{-1}$, $\Omega_{\Lambda}$ = 0.7 and $\Omega_{M}$ = 0.3.

\section{Methods}\label{sec:Methods}

\subsection{Data Reduction}\label{sec:DataReductuon}

\quad \textbf{Calibration:} The HI data were observed as part of a pilot program to observe HI in four low-mass merging systems to determine whether the process of merging affects these systems in a manner similar to their more massive analogues in the VLA B-configuration\footnote{These observations correspond to VLA project 13B-296 and were taken in late 2013.}, and a follow-up program to collect data on an additional six low mass merging systems, as well as VLA C-configuration data for all ten systems\footnote{These observations correspond to VLA project 24A-355 and were taken in mid-2024.}. AGC 727130 was not an object of the observations, but a serendipitous detection, located 7$'$ to the northeast of the targeted interacting dwarf pair. While AGC 727130 has been used as a data point in large statistical studies, there are no specific studies analyzing its resolved properties. We carried out the data reduction using the Common Astronomy Software Application \citep[CASA,][]{casa} using standard calibration routines. Calibration was performed iteratively: we derived an initial solution, flagged radio frequency interference, and recalculated. Target fields were then flagged manually and with the automated CASA task \textit{RGLAG}, which removes statistical outliers. For each final calibrated measurement set, we split off the target field, excluding 1 MHz at either edge of the band (12.5\% of channels on each side) and placed the data on a common visibility-weighting scale using integration-time weights.

\quad \textbf{Imaging:} To produce an optimal 21 cm spectral line cube, we concatenated the VLA-B and VLA-C array data, used the emission-free channels to generate a continuum map (1.7$^{\circ}$ field, 2.0$'$ pixels, 1024 w-planes, two Taylor terms), subtracted the resulting model from the visibilities, and imaged the line data (17$\farcs$1 field, 2$\farcs$0 pixels, 512 w-planes). The line cube was imaged with 62.5 kHz channels and natural weighting, yielding a 6$\farcs$9 $\times$ 6$\farcs$5 beam, and CLEANed blindly to 4.0$\sigma$. A per-channel emission mask was then constructed, and a second cube was imaged with identical parameters but CLEANed to 2.0$\sigma$ within the mask. We smoothed to 7$\farcs$5 resolution, fit and subtracted a per-pixel linear baseline from the emission-free channels, and extracted a 4$\farcm$3$ \times$ 4$\farcm$3 $\times$ 842.5 km s$^{-1}$ sub-cube centered on the optical position and systemic velocity of AGC 727130. The final cube has a velocity resolution of 13.37 km s$^{-1}$, spatial resolution of 1.05 kpc, and an r.m.s. of 0.19 mJy beam$^{-1}$ (0.22 mJy beam$^{-1}$ after primary beam correction), corresponding to a column density sensitivity of 5.9 $\times 10^{19}$ atoms cm$^{-2}$ per channel.

\quad \textbf{Final Determination of Properties:} To construct the total HI distribution, we visually inspected the data cube at multiple spectral and spatial resolutions to identify regions of genuine emission, then generated a channel-by-channel mask to isolate this emission and applied it during the cube collapse to produce the integrated HI intensity (moment 0) map. This procedure ensures that all features in the final map correspond to confidently detected emission while minimizing noise artifacts. From this map we measure a total HI flux of 1.503 $\pm$ 0.188 Jy km s$^{-1}$. The uncertainty was estimated by calculating the number of pixels in the integration region, converting to the number of independent beams, and multiplying by the r.m.s. noise, the channel width, and the square root of the number of independent beams. An integrated HI spectrum, extracted over all detected emission, yields a Gaussian-fit central velocity of 2040.2 $\pm$ 8.0 km s$^{-1}$ and W${20}$ of 103.6 $\pm$ 28.8 km s$^{-1}$. Adopting a luminosity distance of 29.4 Mpc from the systemic velocity, we derive a total HI mass of 3.53 $\pm$ 0.44 $\times$ 10$^{8}$ M${\odot}$. This value is lower than the ALFALFA measurement of 5.37 $\times$ 10$^{8}$ M$_{\odot}$, likely due to beam dilution in Arecibo data, imperfect calibrations between datasets, and the loss of diffuse HI flux from interferometric filtering.

\quad \textbf{Radio Continuum Properties:} In addition to our interferometric HI observations, we observe 1.4 GHz radio continuum over a 128 MHz bandwidth, resulting in radio continuum map with an r.m.s. of 8 $\mu$Jy beam$^{-1}$. This map of 1.4 GHz radio continuum flux density, in the absence of an active galactic nuclei, can be used to trace star-formation in an extragalactic source \citep{yun01, bell03, murphy11}. However, the region coincident with both the optical and HI components of AGC 727130 show no statistically significant radio continuum detection. Using our r.m.s. value of 8 $\mu$Jy beam$^{-1}$, a luminosity distance 29.4 Mpc, a statistical significance threshold of 3$\sigma$, and using Equation 17 in \citet{murphy11}, we derive the 3$\sigma$ upper limit on the star-formation rate of AGC 727130 as 1.2 $\times$ 10$^{-3}$ $M_{\odot}$ yr$^{-1}$. This result is in agreement with spectroscopic results from \citet{Suess16} who, as part of a follow-up on a subset of possible OH detections, observe no optical emission lines, including H$\alpha$, in an Apache Point Observatory spectrum of the source. 

\quad The ALFALFA–SDSS catalog \citep{Durbala20} reports a stellar mass of 10$^{7.85}$ M$_{\odot}$ for AGC 727130, derived from Sloan Digital Sky Survey \citep[SDSS;][]{York00} photometry, along with an absolute $i$-band magnitude of –16.22 and no detectable star formation, despite the galaxy’s distinctly blue color. These values correspond to an $i$-band mass-to-light ratio of 0.35. Based on its stellar mass, AGC 727130 would be expected to form stars at a rate close to our detection limit, consistent with the gas-rich main sequence of star-forming galaxies described by \citet{McGaugh17}. Instead, the galaxy appears to be forming fewer stars than expected, although it is not clearly quenched. Its blue color further suggests that, if quenching has occurred, the event must have been relatively recent.

\subsection{Defining the Cosmic Web}\label{sec:WebDefine}

\quad To define the large-scale filamentary network that constitutes the cosmic web, we use the Discrete Persistent Structure Extractor \citep[DisPerSE;][]{disperse1, disperse2}. DisPerSE is a topological, scale-free algorithm that returns the extrema of various manifolds, and can be used to identify filamentary structures in a three dimensional dataset. For the purposes of identification of the large-scale structure of the universe, this dataset consists of the RA, Dec, and Z for galaxies in a given patch of sky. From these points, DisPerSE returns a set of critical points that can be connected to trace the filamentary network present within the data. The usage of DisPerSE on astronomical databases for this purposes has been used extensively in the recent past with studies ranging from large-area low redshift surveys \citep{Odekon18}, to small fields of view at high redshift \citep{Kraljic18}, to large-scale cosmological simulations \citep{Hasan23}, to studies comparing the effect of the cosmic web in both simulations and observations \citep{Luber25}.

\quad For our application of DisPerSE, we use a galaxy catalog with optical positions (RA and DEC) and high confidence spectroscopic redshifts from SDSS. Specifically, we queried SDSS for all galaxies with spectroscopic redshifts in the range 0.0 and 0.075, and all galaxies with an RA and DEC within a 40$^{\circ}$ $x$ 40$^{\circ}$ box centered on AGC 727130. This allows us to, at the redshift of AGC 727130, probe a 20 Mpc $x$ 20 Mpc area of AGC 727130 which is suitable for determining large-scale filaments. In our application, we used the mirror boundary condition, and a critical point significance threshold of six. This ensures that we analyze only the features most truly likely to be physical filamentary structures. 

\begin{figure*}
\begin{center}
\includegraphics[width=\textwidth]{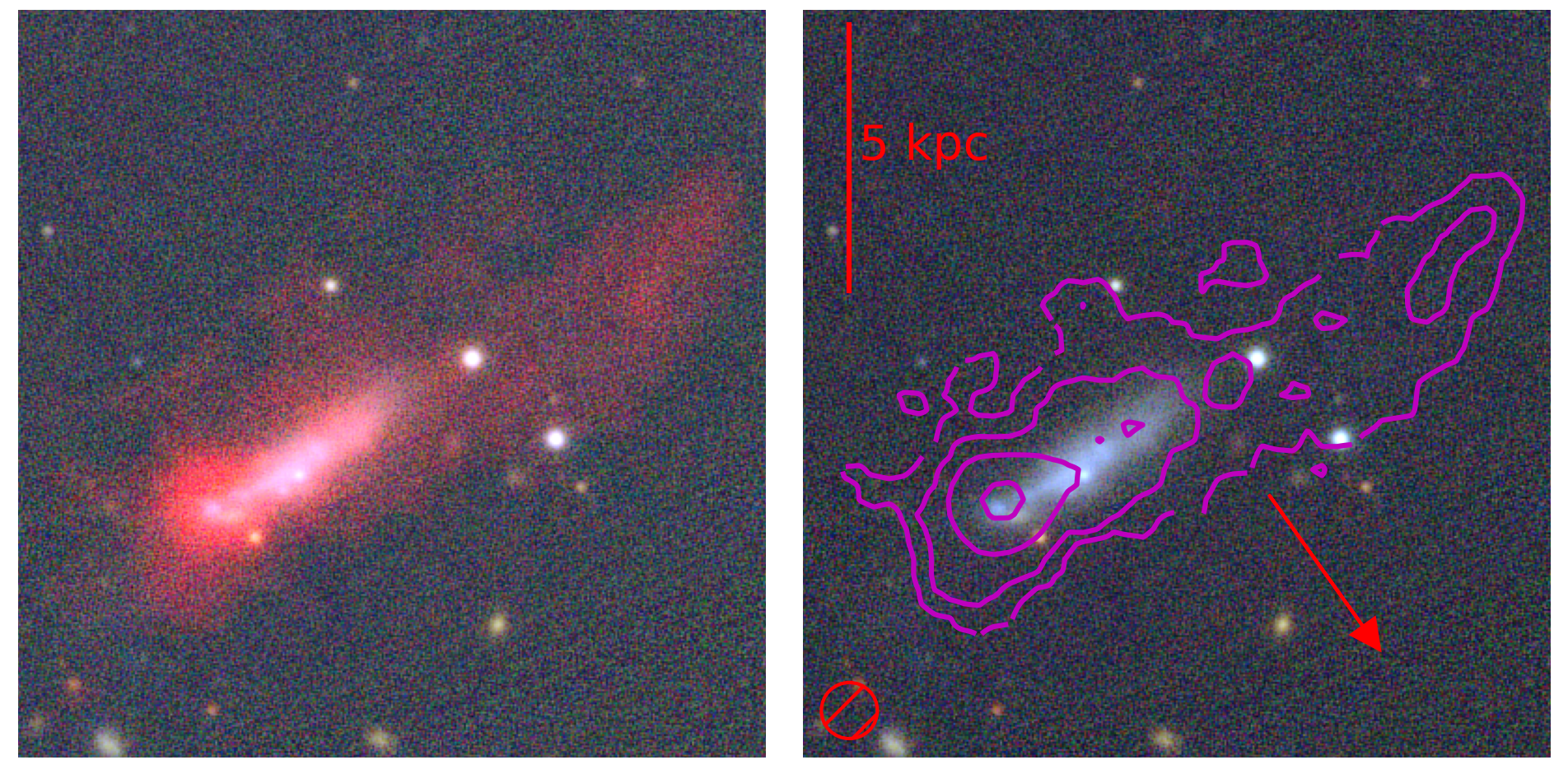}
\caption{\textbf{\textit{Left:}} False-color RGB composite image of AGC727130, created using Pan-STARRS g-, r-, i- band images \citep{panstarrs}, with a red-scale overlay showing the spatial distribution of the neutral atomic gas. This image is oriented with north pointing upwards, and west pointing rightwards. \textbf{\textit{Right:}} The same optical image, with HI column density contours overlaid with levels 2$^{n}$ $\times$ 10$^{20}$ atoms cm$^{-2}$ for n = 1, 2, 3, 4. The hatched circle in the bottom left corresponds to the restoring beam for the HI observations, the red scale bar in the top left corresponds to 5 kpc in projection, and the arrow in the bottom right points to the direction of the interacting pair 60 kpc away in projection.}\label{fig:Mom0+Opt_RBG}
\end{center}
\end{figure*}

\section{Results}\label{sec:Results}

\subsection{Neutral Gas Morphology and Kinematics}

\begin{figure}
\begin{center}
\includegraphics[width=\columnwidth]{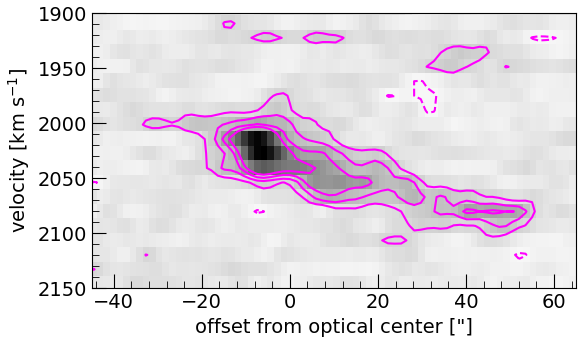}
\caption{A position-velocity diagram taken across the major axis of the galaxy. The x-axis corresponds to the angular offset along the axis from optical center, the y-axis velocity corresponds to the observed line-of-sight velocity, and the greyscale and contours indicate the intensity level.}\label{fig:PVSlice}
\end{center}
\end{figure}

\quad To understand the processes shaping AGC 727130, we examine its neutral hydrogen morphology and kinematics. In Figure~\ref{fig:Mom0+Opt_RBG}a, we display the distribution of HI in redscale, overlaid on a color RGB image of AGC 727130 created from PanSTARRS photometric data, and in Figure~\ref{fig:Mom0+Opt_RBG}b, we show the HI distribution as column density contours overlaid on the color RGB image of AGC 727130. The figure clearly reveals a strong asymmetry in the HI gas (at the column density level of $\sim$ 5 $\times$ 10$^{20}$ atoms cm$^{-2}$) compared to the galaxy’s relatively typical stellar component (down to the $\sim$ 24.2 mag arcsec$^{-2}$ in the g-band at the 1$\sigma$ level). In particular, a significant portion of the neutral hydrogen appears to be displaced from the stellar disk toward the northwest, following the galaxy’s major axis. In Figure~\ref{fig:PVSlice}, we present a position-velocity (PV) slice extracted along the major axis of AGC 727130. The diagram further emphasizes the pronounced asymmetry seen in the total intensity map, showing a notable extension of gas offset toward the southern region relative to the optical center. Additionally, the PV slice reveals that the velocity of the gas increases progressively as one moves westward from the optical center, rather than reaching a flat rotation curve typical of undisturbed galaxies \citep[e.g.][]{Swaters09}. This rising velocity gradient provides evidence that gas is being actively removed or stripped from the galaxy. These spatial and kinematic features suggest that external forces are affecting the galaxy, likely through environmental interactions, which we explore in the following analysis.

\subsection{Gravitational Influence of Neighboring Dwarfs}

\quad The first, and perhaps most straightforward, suspect for causing the asymmetry is a gravitational interaction. AGC 727130 is more than 1.5 Mpc away from a more massive neighbor (M$_{*}$ $>$ 10$^{9.5}$ M$_{\odot}$) but the galaxy does lie in relatively close proximity to a pair of interacting dwarf galaxies \citep{Stierwalt15}. 

\quad For our subsequent analysis, we adopt a redshift of 0.00683 for AGC 727130, and a redshift of 0.00673 for the nearby interacting pair, as measured by their HI systemic velocity, respectively. The redshift for the nearby interacting pair is in agreement with reported values from SDSS, but we note that the SDSS redshift for AGC 727130 is significantly different, 0.004, and speculate that this is due to the lack of significant optical emission lines which causes redshift determination to be challenging. With our adopted HI redshifts, AGC 727130 has a projected separation of 60 kpc, found adopting a redshift of 0.00683 using the measured HI systemic velocities.

\quad To evaluate whether the nearby dwarf galaxy pair could exert a significant tidal influence on AGC 727130, we estimate the tidal radius, $r_{t}$. The tidal radius places an upper limit on the size of AGC 727130 beyond which an external perturber, in this case the dwarf pair, will have a stronger gravitational influence on the galaxy's gas or stars than the gravitational pull of the galaxy itself. We calculate $r_{t}$ using the standard approximation:
\begin{equation}
    r_{t} \approx D \left( \frac{M_{gal}}{3M} \right)^{\frac{1}{3}}
\end{equation}
where D is the distance between AGC 727130 and the center of the dwarf pair, $M_{gal}$ is the combined gas and stellar mass of AGC 727130, and M is the combined gas and stellar mass of the dwarf pair \citep{BinneyTremaine}. For this scenario, we adopt $M_{gal} = M_{*} + M_{HI}$, with $M_{*} = 10^{7.85} M_{\odot}$, as reported in \citet{Durbala20}, and $M_{HI} = 10^{8.55} M_{\odot}$, as calculated in Section~\ref{sec:DataReductuon}, for a total mass of $M_{gal} = 10^{8.62} M_{\odot}$. For the dwarf pair, using the same definition modified to be the sum of two systems $M = M_{*,1} + M_{*,2} + M_{HI,1} + M_{HI,2}$, with $M_{*,1} = 10^{7.57} M_{\odot}$ and $M_{*,2} = 10^{7.70} M_{\odot}$ as reported in \citep{Mendel14}, and $M_{HI,1} = 10^{8.94} M_{\odot}$ and $M_{HI,2} = 10^{8.07} M_{\odot}$ using the VLA data described in Section~\ref{sec:DataReductuon}, for a total mass of $M = 10^{9.03} M_{\odot}$. We take to be the distance of AGC 727130 to the center of gravity of the interacting nearby pair, D, to be the projected distance between them of 60 kpc. We do not include dark matter in this calculation as the total masses are within a factor two of each other, and are likely to have similar dark matter halo masses within reasonable error estimates. This leads to a value of 30 kpc for the tidal radius of AGC 727130 for this interaction.

\quad Comparing this to the optical radius for AGC 727130, 3.1 kpc, found by fitting an ellipse to the PanSTARRS optical g-band image, and the HI radius, 6.33 kpc, found by fitting an ellipse to our total HI emission map, we find that these radii are well below the tidal radius. Rotation curves for dwarf galaxies tend to flatten after roughly two disk scale lengths \citep{Swaters09} suggesting the dark matter halo for AGC 727130 is also likely to be well within this tidal radius. This indicates that AGC 727130 should remain gravitationally bound and largely unaffected by tidal stripping at the current separation. 

\quad Although a present-day tidal interaction is unlikely, we also test whether a past close encounter could explain the observed feature by comparing the dynamical timescale of tail formation with the system’s crossing time. To calculate the dynamical timescale of the tail, $t_{D}$ we use the following equation presented in \citet{Schweizer82}:
\begin{equation}
    t_{D} \approx \frac{0.8r_{p}\cot(\beta)}{v}
\end{equation} 
where $r_{p}$ is the projected radius from the optical center to the tip of the tail, $v$ is the line-of-sight velocity difference of the tip of the tail to the center of the optical galaxy, and $\beta$ is the angle between the radius vector of the galaxy and the line-of-sight. For AGC 727130, $r_{p}$ = 8.97 kpc, and $v$ = 53.41 km s$^{-1}$. The value for $\beta$ is somewhat challenging to constrain, as determining an inclination for this galaxy is not straightforward, but we adopt a value of 60$^{\circ}$ as a reasonable estimate. These values lead to a dynamical timescale of the tail formation as $t_{D} \approx$ 75.9 Myr. To calculate the crossing time, $t_{C}$, we use the following expression:
\begin{equation}
    t_{C} \approx \frac{D}{\sqrt{3}\Delta v_{LOS}}
\end{equation} 
where $\Delta$ $v_{LOS}$ is the difference of line-of-sight velocities between AGC 727130 and the interacting pair and D is the distance between AGC 727130 and the interacting pair. For this scenario, we have $\Delta$ $v_{LOS}$ $=29.7$ $km$ $s^{-1}$, measured from the difference in HI systematic velocities, and D = 59.4 kpc, and find a crossing time of 1130.7 Myr.

\quad The above calculations show that the crossing time for this system is a factor 15 greater than the dynamical timescale of the HI tail, if it was caused by a previous interaction. Additionally, as indicated in Figure~\ref{fig:Mom0+Opt_RBG}b, the HI extension is not oriented in the direction of the dwarf pair, which may be expected if the gas distribution were tracing some previous interaction. Moreover, if there was an impactful interaction with the nearby dwarf pair, we would likely expect some star-formation enhancement \citep{Stierwalt15}, as opposed to the non-detection of star-formation. Lastly, at the level of 3 $\times 10^{20}$ atoms cm$^{-2}$, AGC 727130 is more than 50 kpc away from the shared gas envelope of the interacting pair further suggesting that it has little influence on the current morphology of AGC 727130. Taken all together, given the current separation, mass ratio, and HI distribution, these results motivate the consideration of alternative explanations for the HI asymmetry.

\begin{figure}
\begin{center}
\includegraphics[width=0.95\columnwidth]{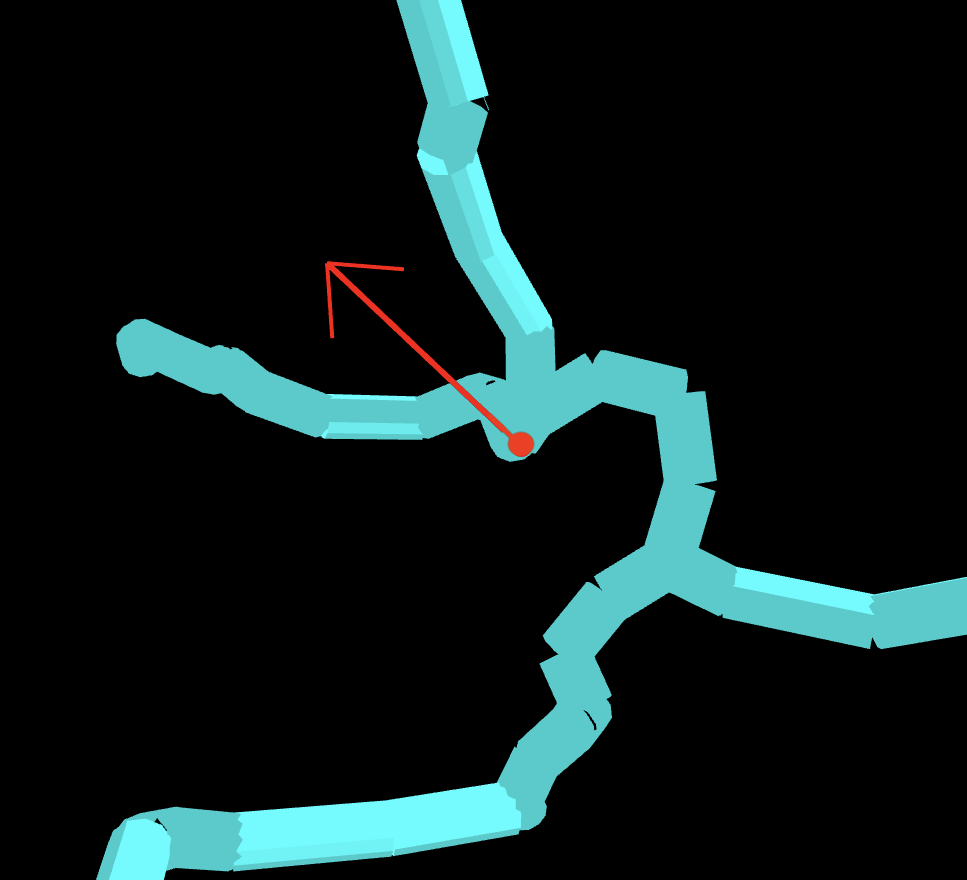}
\caption{The orientation of AGC 727130 within the cosmic web. The blue cylinders indicate the spine of the filamentary network, as outputted by DisPerSE, and their width corresponds to 1 Mpc. The red dot marks the position of AGC 727130, and we note that the galaxy is placed within the intersection of the filaments at this point. The arrow indicates the direction of the asymmetric HI extension, corresponding to pointing towards the northwest in the plane-of-the-sky as seen in Figure~\ref{fig:Mom0+Opt_RBG}.}\label{fig:LSS}
\end{center}
\end{figure}

\subsection{Cosmic Web Ram Pressure Stripping}

\begin{figure}[t!]
\begin{center}
\includegraphics[width=0.95\columnwidth]{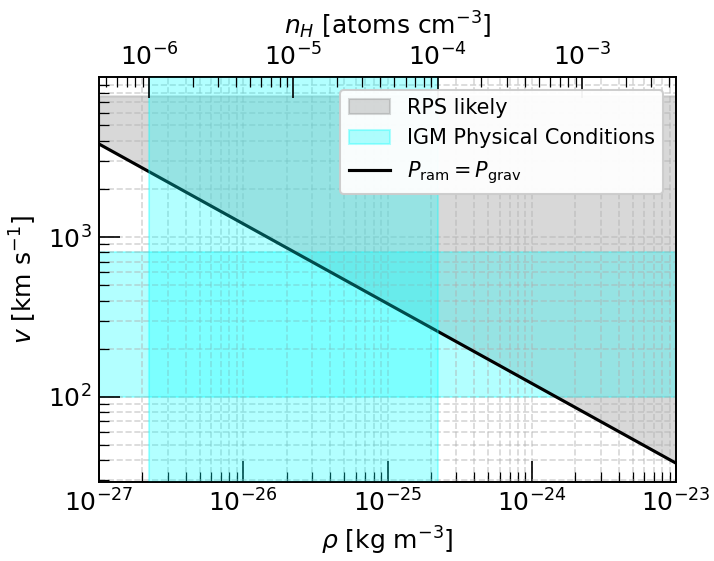}
\caption{Ram pressure stripping criterion for AGC 727130. The grey-shaded region indicates combinations of intergalactic medium density and galaxy velocity where the external ram pressure exceeds the galaxy’s restoring force, making gas removal via ram-pressure stripping likely. The cyan regions correspond to the regions of parameter space expected in cosmic web filaments.}\label{fig:RPS}
\end{center}
\end{figure}

\quad The next possibility we consider is ram pressure stripping. Although RPS often manifests as a truncation of the HI disk relative to the optical disk, such signatures can be difficult to detect because of projection effects and the orientation of the galaxy with respect to its motion through the surrounding medium. Additional evidence comes from the simulations of \citet{Kronberger08}, who investigated the internal gas dynamics of galaxies undergoing RPS. They find that for galaxies angled edge-on to the stripping medium, the kinematic and optical centers become offset, and the outer rotation curves decline rather than flatten. Both of these features are observed in AGC 727130 (see Figure~\ref{fig:RPS}), strongly suggesting that RPS is the origin of its disturbed HI morphology and kinematics.

\quad To further assess this scenario, we examine the large-scale structure environment of AGC 727130 to determine whether a medium capable of stripping its cold neutral gas is present. We use the Discrete Persistent Structure Extractor \citep[DisPerSE;][]{disperse1,disperse2} to identify filamentary structures in the SDSS spectroscopic catalog. Details of this procedure are given in Section~\ref{sec:WebDefine}. As shown in Figure~\ref{fig:LSS}, AGC 727130 lies at the intersection of several filaments, an environment particularly well suited to hosting diffuse, shock-heated gas. These junctions are predicted by simulations to accumulate baryons, including hot, ionized gas, as matter flows along filaments toward higher-density regions. Such a medium could provide the conditions necessary for ram pressure stripping, even in the absence of a massive group or cluster halo. Observational support for this scenario comes from the proof-of-concept DisPerSE study \citep{disperse2}, which identified a filament intersection in galaxy redshift space and subsequently detected an associated reservoir of hot gas in X-ray observations. The case of AGC 727130 may therefore represent a lower-mass analog, where gas stripping is driven by interaction with a filament-fed, warm-hot intergalactic medium.

\section{Discussion}\label{sec:Discussion}

\quad AGC 727130 combines the stellar distribution of a dwarf irregular with an offset, asymmetric HI reservoir and a blue color despite its lack of star formation, pointing to an environmentally driven transformation. The selective removal of the gas component, particularly along a singular axis and without clear evidence of stellar disruption, points to a mechanism such as RPS. However, the galaxy is not embedded in a galaxy cluster or massive group, where RPS is typically activated by the intracluster media. Instead, our analysis using DisPerSE reveals that AGC 727130 lies at the intersection of several large-scale filaments of the cosmic web. Simulations predict that filamentary junctions accumulate baryons, especially hot ionized gas, as matter flows along filaments into higher-density nodes. \citep{Martizzi19,Gouin22,Espinosa24}. The diffuse hot ionized gas is difficult to detect observationally \citep{Tanimura22} but could be sufficient to exert a diffuse ram pressure on low-mass galaxies over gigayear timescales, given the nature of filaments as diffuse, large-scale structures.

\quad Observational precedents for this scenario exist: the original DisPerSE proof-of-concept study by \citep{disperse2} detected an X-ray-bright hot gas reservoir at a filamentary intersection, independent of any massive halo. Similarly, the case of AGC 727130 may represent another instance of this phenomenon, where the local geometry of the cosmic web, rather than the influence of a massive host, facilitates gas removal. Recent simulations have shown that cosmic web stripping is particularly efficient for dwarf galaxies, where shallow potential wells and extended HI disks make them highly susceptible to mild but persistent external pressures \citep{BenitezLlambay13, Thompson23, Pasha23, Benavides25}.

\quad To assess whether ram-pressure stripping could be responsible for the gas depletion in AGC 727130, we compare the expected ram pressure from the cosmic web to the galaxy’s internal restoring force. To do this we use the formalism in \citet{GunnGott}, where by combining Equations 61 and 62 we get the following:
\begin{equation}
    \rho v^{2} = 2 \pi G \Sigma_{*} \Sigma_{g}
\end{equation}\label{eq:rps}
where $\rho$ is the gas density within the filaments, $v$ is the velocity of the galaxy with respect to the filament, and $\Sigma_{*}$ and $\Sigma_{g}$ are the surface star and gas densities of the galaxy at a given radius.
Figure~\ref{fig:RPS} shows the critical boundary in density–velocity space where the external ram pressure equals the gravitational restoring force at the stellar and gaseous surface densities of 2.5 and 3.2 M$_{\odot}$ pc$^{-2}$, respectively. Above this line, the ram pressure exceeds the binding force on the galaxy’s gas, and stripping is expected to occur. Below the line, the galaxy can retain its gas despite the external medium. The top axis converts the physical density into hydrogen number density assuming a primordial composition. This framework illustrates the environmental conditions required for effective RPS and provides a direct comparison between the expected cosmic web gas densities and galaxy velocities within the cosmic web.

\quad We note that the RPS criterion form \citet{GunnGott} used here provides a lower limit on the required ram pressure, as it neglects the contribution of the dark matter halo to the restoring force. Dwarf galaxies like AGC 727130 are typically dark matter dominated \citep{Bullock17}, and even a modest baryon-to-dark matter mass ratio of 0.1 in the galaxy center \citep{Mori00} can increase the restoring force, and thus the threshold for complete stripping, by up to an order of magnitude. However, AGC 727130 is not completely stripped: it retains $\sim$50\% of its HI within the stellar disk, and the dense gas morphology on the wind-leading side remains largely unperturbed. In this partial stripping regime, the effective threshold depends sensitively on the gas and dark matter density profiles \citep{McCarthy08}. Simulations of comparable dwarf galaxies find that the ram pressure required to strip $\sim$50\% of the gas mass is typically only $\sim$0.5 dex lower than that required for full stripping \citep{Fillingham16,Zhu24}. Thus, while Equation~\ref{eq:rps} may slightly underestimate the pressure required to reproduce the observed partial stripping in AGC 727130, the discrepancy is likely modest compared to the plausible range of filament densities and velocities.

\quad These results suggest that AGC 727130, despite not residing in a massive cluster environment, may still be susceptible to ram pressure stripping due to its position at the intersection of three filaments. Both simulations and observations indicate that filament intersections often exhibit elevated gas densities of 10$^{-6}$ $<$ $n_{H}$ $<$ 10$^{-4}$ cc, and typical galaxy velocities along filaments, particularly at the intersection of filaments, are of order a few hundred km s$^{-1}$, similar to the velocity dispersion of galaxies in groups \citep{Smith12, Cautun14, Tempel14, Martizzi19} This is illustrated in Figure~\ref{fig:RPS} with the cyan regions indicating the range of possible values for each parameter. Combined with the relatively modest restoring force of a low-mass galaxy like AGC 727130, such enhanced filamentary environments could produce ram pressures approaching or exceeding the stripping threshold, and in Figure~\ref{fig:RPS}, we demonstrate this by showing that there is an overlap between the expected properties of cosmic web filaments (cyan regions) and the ram pressure required to strip gas from the galaxy. Additionally, local variations in ISM and cosmic web gas density can increase the ram pressure experienced by the galaxy which would further increase the likelihood, and amount of, gas that is stripped from the disk \citep{Quilis00,Tonnesen08,Tonnesen09}. Thus, the intersection of multiple filaments may represent an intermediate regime of environmental processing, denser than the general cosmic web but less extreme than rich clusters, where galaxies like AGC 727130 can lose gas via ram pressure.

\quad If AGC 727130 is indeed being stripped by the ambient medium of the cosmic web, this has broader implications for our understanding of low-mass galaxy evolution. If overdensities in the large-scale structure can remove gas from dwarfs, then some dark matter halos predicted by simulations may host galaxies that were stripped and quenched before becoming observable in optical or HI surveys. A natural question raised by this interpretation is why quenched dwarfs are not more common in the field if ram pressure stripping can operate efficiently in filaments. One possibility is that filamentary stripping acts more slowly than cluster-driven processes, removing gas over gigayear timescales without always producing rapid quenching. This may result in disturbed HI morphologies, like in AGC 727130, without yielding a large quenched population. Moreover, only a subset of dwarfs, those with shallow potential wells or located at dense filamentary junctions, may be especially vulnerable. Finally, quenched field dwarfs are expected to be optically faint and difficult to identify, potentially biasing observed samples. Together, these effects may reconcile the apparent rarity of quenched field dwarfs with the physical plausibility of cosmic web ram pressure stripping.

\quad Testing this hypothesis motivates several avenues for future work. Deeper observations of the environment of AGC 727130, particularly in X-ray mapping of hot gas, could directly probe the presence of a surrounding medium. Statistical studies of low-mass galaxies at filament intersections can assess whether disturbed HI morphologies are more common in such environments. Higher-resolution hydrodynamical simulations that include diffuse filamentary gas and realistic dwarf orbits are also needed to test whether observed asymmetries can be reproduced. More speculatively, cosmic web stripping may set an observational floor for galaxy formation, below which halos cannot retain their gas even if they can form stars initially. As wide-field HI and optical surveys push to lower masses and higher sensitivities, identifying more systems like AGC 727130 will be essential for constraining this threshold and for understanding how the cosmic web regulates galaxy evolution.

\begin{acknowledgments}

\quad We thankfully acknowledge the helpful comments and suggestions from the anonymous referee that added to the quantification of results and scientific context presented in this work.

\quad The National Radio Astronomy Observatory and Green Bank Observatory are 
facilities of the U.S. National Science Foundation operated under 
cooperative agreement by Associated Universities, Inc.

\end{acknowledgments}

\bibliography{references}{}
\bibliographystyle{aasjournalv7}

\end{document}